\def\preprint{1}			

\ifdefined\preprint
  \documentclass[preprint,review,12pt]{elsarticle}
\fi
\ifdefined\wordcount
  \documentclass[final,3p,times,twocolumn]{elsarticle}
\fi
\ifdefined\final
  \documentclass[final,3p,times,twocolumn]{elsarticle}
\fi

\usepackage{graphicx,stfloats}
\usepackage{color}

\usepackage{bm}
\usepackage{amsmath}
\usepackage{amstext}
\usepackage{epstopdf}
\usepackage{booktabs}
\usepackage{textcomp}

\graphicspath{{Fig/}}

\biboptions{sort&compress}

\begin{document}

\begin{frontmatter}

\title{Modeling pressure effects on the turbulent burning velocity for lean hydrogen/air premixed combustion}

\author[fir,sec]{Zhen Lu}

\author[fir,sec,thi]{Yue Yang}
\ead{yyg@pku.edu.cn}

\address[fir]{State Key Laboratory for Turbulent and Complex Systems, College of Engineering, Peking University, Beijing 100871, China}
\address[sec]{BIC-ESAT, Peking University, Beijing 100871, China}
\address[thi]{CAPT, College of Engineering, Peking University, Beijing 100871, China}
\cortext[cor1]{Corresponding author:}

\begin{abstract}

We investigate and model pressure effects on the turbulent burning velocity over a wide range of pressures and turbulence intensities with the direct numerical simulation (DNS) of statistically planar turbulent premixed flames for lean hydrogen/air mixture.
DNS results indicate that the stretch factor has an impact on the turbulent burning velocity and flame surface area at elevated pressures. In particular, the enhanced stretch factor at high pressures increases the ratio of turbulent and laminar burning velocities, diminishing the ``bending'' effect.
Based on a good consistency between turbulent and laminar burning velocities with respect to flame stretch, a lookup table formed by laminar flame data is employed to model the stretch factor in turbulent flames at various pressures. A predictive model for the turbulent burning velocity is then developed by combining sub models of the stretch factor and flame surface area. The overall good agreement between model predictions and DNS results demonstrates that the proposed model is able to quantitatively predict the turbulent burning velocity over a wide range of pressures and turbulent intensities in homogeneous isotropic turbulence.

\end{abstract}


\end{frontmatter}

\ifdefined \wordcount
\clearpage
\fi

\section{Introduction}
\label{sec:intro}

The turbulent burning velocity $s_T$ is one of the most important quantities characterizing turbulent premixed combustion~\cite{Peters_2000,Lipatnikov_2002,Driscoll_2008}. Extensive efforts have been made to investigate the dependence of $s_T$ on various factors such as the fuel composition and chemistry, turbulence intensity, and pressure.

Considering engine relevant conditions, the pressure influence on $s_T$ has been widely reported in experimental works~\cite{Daniele_2011,Kobayashi_2013,Bradley_2013,Venkateswaran_2015,Fragner_2015,Wang_2015,Nguyen_2019}. Most studies noticed that the laminar burning velocity  $s_L^0$ of unstretched flames decreases with pressure, whereas $s_T$ is insensitive to or even increased with pressure~\cite{Venkateswaran_2015}, so the ratio $s_T/s_L^0$ rises with pressure. Another important observation is the suppression of the ``bending'' effect~\cite{Lipatnikov_2002} of $s_T/s_L^0$ at elevated pressures~\cite{Daniele_2011,Kobayashi_2013}. These phenomena have been attributed to small-scale wrinkling of turbulent flame surfaces~\cite{Kobayashi_2002_B,Kobayashi_2013}.
Various power laws and scaling methods~\cite{Venkateswaran_2015,Nguyen_2019} have been proposed to collapse $s_T$ curves with respect to pressure, but the pressure effects on $s_T$ has not been clearly elucidated and quantitatively modeled.

The direct numerical simulation (DNS) is useful to interrogate detailed information of turbulent combustion~\cite{Im_2016}. Eulerian \cite{Day_2009,Aspden_2017} and Lagrangian \cite{Uranakara_2016} investigations have been employed to explain interactions between turbulence and local flame speed, and to develop correlations between flame stretch and $s_T$~\cite{Chen_1998}.
The bending curve has been reproduced by DNS~\cite{Nivarti_2017,You_2019} at 1 atm, confirming it is caused by the inhibited growth of flame areas.

In recent years, three-dimensional DNS of turbulent premixed flames at elevated pressures were reported by several groups for practical interest~\cite{Savard_2017,Wang_2017,Wang_2018,Savard_2019}.
Savard \emph{et al}.~\cite{Savard_2017} studied pressure effects on complex fuels in engine relevant conditions. By comparing turbulent flame statistics at 1 and 20 atm with the same Karlovitz number, they argued that the flame area, rather than the stretch factor, causes differences in $s_T$ for iso-octane. Wang \emph{et al}.~\cite{Wang_2017} reported a series of DNS of lean methane/air turbulent flames with various turbulence intensities at 20 atm, and also found that the flame area is the dominant factor for $s_T$. By contrast, some experimental studies~\cite{Venkateswaran_2011,Venkateswaran_2015} indicated that the stretch sensitivity could be crucial for mixtures with negative Markstein numbers. 

A predictive model of $s_T$ is of practical interest for industrial design and combustion modeling~\cite{Peters_2000,Vervisch_1995}. A number of empirical models have been proposed for $s_T$~\cite{Lipatnikov_2002,Driscoll_2008}. However, most of them are not validated over the wide range of pressures and turbulence intensities, and the various power-law scalings strongly depend on empirical parameters~\cite{Lipatnikov_2002,Driscoll_2008}.
Recently, You and Yang~\cite{You_2019} proposed a $s_T$ model based on Lagrangian statistics of propagating surfaces~\cite{Girimaji_1992,Zheng_2017}. By estimating the flame area with universal model constants obtained from homogeneous isotropic turbulence (HIT), the model successfully predicts $s_T/s_L^0$ for various fuels at 1 atm, but the constant $s_L^0$ used in the modeled area growth rate of flames limits its application to standard pressure.

In this work, we carry out a systematic DNS study of lean H$_2$/air turbulent premixed flames at a wide range of pressures and turbulence intensities.
Statistics of $s_T$ and flame surface areas are investigated to gain insight of underlying governing processes for the bending of $s_T$.
Effects of the flame stretch on local flame propagation in turbulent flames are then analyzed thoroughly over the wide range of parameters.
According to the analysis, we propose a new model for predicting $s_T$, considering variations of both the stretch factor and flame area with respect to pressure.
Finally, the model is validated by DNS results.

\section{Simulation overview}
\subsection{DNS parameters}
\label{sec:parameters}

For the DNS of turbulent premixed flames, we consider the free propagation of a statistical planar premixed flame along the streamwise direction in statistically stationary HIT at a range of pressures $p=$ 1, 2, 5, 10 atm.
The unburnt gas is a lean hydorgen/air mixture with the equivalence ratio $0.6$ at the temperature $T_u=300\;\mathrm{K}$.
For each pressure, the thermal thickness $\delta_L^0 = (T_b-T_u)/\lvert\nabla T\rvert_\mathrm{max}$, the laminar flame speed $s_L^0$, the displacement speed $s_d^0$ at the temperature $T_\mathrm{peak}^\mathrm{F}$ corresponding to the peak fuel consumption rate obtained in the freely propagating laminar flame,  and the flame Reynolds number $\mathrm{Re_F} = s_L^0\delta_L^0/\nu$ with the kinematic viscosity $\nu$ are listed in Table~\ref{tab:laminar}, where $T_b$ is the temperature of the burnt gas, $\lvert\nabla T\rvert_\mathrm{max}$ denotes the maximum temperature gradient in the laminar flame, and the superscript $0$ denotes a quantity in unstretched flames.

\begin{table}[htbp]
  \centering
  \caption{Parameters of unstretched laminar flames}
  \label{tab:laminar}
  \begin{tabular}{ccccc}
    \toprule[1.5pt]
    $p\;(\mathrm{atm})$                         & 1       & 2       & 5       & 10      \\
    \midrule[1.0pt]
    $\delta_L^0\;(\mathrm{\mu m})$              & 365.4   & 178.8   & 83.16   & 58.40   \\
    $s_L^0\;(\mathrm{m/s})$                     & 0.833   & 0.646   & 0.410   & 0.240   \\
    $s_d^0\;(\mathrm{m/s})$                     & 3.22    & 2.66    & 1.81    & 1.12    \\
    $T_\mathrm{peak}^\mathrm{F}\;(\mathrm{K})$  & 1294.0  & 1366.4  & 1471.3  & 1569.2  \\
    $\mathrm{Re}_\mathrm{F}$                    & 15.7    & 11.9    & 8.77    & 7.23    \\
    \bottomrule[1.5pt]
  \end{tabular}
\end{table}

For each pressure, we conduct four DNS cases with a range of turbulence intensities $u'/s_L^0 = 2$, 5, 10, and 20 (i.e., in total of 16 DNS cases in the present study).
The parameters for combustion DNS are listed in Table~\ref{tab:turbulent}, where $u'$ is the rms velocity fluctuation, and the ratio between turbulence integral scale $l_t$ and flame thickness $\delta_L^0$ is kept as unity. Dimensionless numbers are listed in Table~\ref{tab:turbulent}.
The Damk\"ohler number $\mathrm{Da} = \tau_e/\tau_f = (l_t/\delta_L^0)(s_L^0/u')$ is defined as the ratio of the integral timescale $\tau_e$ and the flame timescale $\tau_f$.
The Karlovitz number $\mathrm{Ka} = (u'/s_L^0)^{\frac{3}{2}}(\delta_L^0/l_t)^{\frac{1}{2}}$ is defined with the dissipation rate~\cite{Aspden_2017}.
The turbulence Reynolds number $\mathrm{Re}=\mathrm{Re}_0\mathrm{Re_F}$ is linked with $\mathrm{Re}_0=\left(u'/s_L^0\right)\left(l_t/\delta_L^0\right)$ in the regime diagram.
We remark that the definitions of $\mathrm{Da}$, $\mathrm{Ka}$, and $\mathrm{Re}_0$ are independent of pressure, so the DNS cases with the same $u'/s_L^0$ but at different pressures are at the same point in the regime diagram~\cite{Peters_2000}.

\begin{table}[htbp]
	\centering
  \caption{Parameters of turbulent combustion DNS}
  \label{tab:turbulent}
	\begin{tabular}{cccccc}
		\toprule[1.5pt]
		$ u^\prime / s_L^0 $  & 2     & 5     & 10    & 20    \\
		\midrule[1.0pt]
    $ l_t / \delta_L^0 $  & 1     & 1     & 1     & 1     \\
    $ \mathrm{Re}_0 $     & 2     & 5     & 10    & 20    \\
		$ \mathrm{Da} $       & 0.5   & 0.2   & 0.1   & 0.05  \\
		$ \mathrm{Ka} $       & 2.828 & 11.18 & 31.62 & 89.44 \\
		\bottomrule[1.5pt]
	\end{tabular}
\end{table}

\subsection{Numerical methods}
\label{sec:numerics}

The present DNS solves the low Mach number, variable density formulation of transport equations for mass, momentum, species, and temperature using the NGA code~\cite{Desjardins_2008}, with the Strang splitting applied for transport--chemistry coupling~\cite{Ren_2008}. For the transport part, equations are advanced by an iterative semi-implicit Crank--Nicolson scheme~\cite{Pierce_2001}. A second-order centered, kinetic-energy conservative finite difference scheme is used for discretizing spatial derivatives in momentum equations, and a third-order bounded QUICK scheme~\cite{Herrmann_2006} is used for treating convection terms in scalar transport equations of species mass fractions and temperature. The time integration of chemical substep is performed by the stiff solver DVODE~\cite{Brown_1989}. The detailed nine-species H$_2$/air mechanism \cite{Li_2004} is employed, and molecular transport is modeled with constant Lewis numbers for each species~\cite{Burali_2016}. Each DNS case is first run for at least $10\tau_e$ to reach a statistically stationary state, and then statistics are calculated over a period of at least 15$\tau_e$.

The computational domain is a cuboid with sides $ L_x \times L_y \times L_z = 12 L \times L \times L$ and $ L = 5.3 l_t $. This domain is discretized on uniform grid points $ N_x \times N_y \times N_z = 12 N \times N \times N $. The numerical resolution in all the cases is ensured to resolve the smallest turbulent and flame length scales by the criterion $k_{\mathrm{max}}\eta \geq 1.5$~\cite{Pope_2000} and a minimum of 24 grid points within a flame thickness $\delta_L^0$, respectively, where $ k_{\mathrm{max}} = \pi N/L$ is the maximum wavenumber magnitude in DNS of HIT and $\eta$ is the Kolmogorov length scale. To meet the criteria, we set $N=128$ for all the cases listed in Table~\ref{tab:turbulent} except $N=256$ for the case with $u'/s_L^0=20$ and $p= 1$ atm.
The timestep is controlled by the CFL number less than 0.5.
The computational domain has inflow and outflow conditions in the streamwise $x$-direction, and periodic boundary conditions are imposed in lateral $y$- and $z$-directions. The inflow is generated by a separate DNS of non-reacting, statistically stationary HIT, and it is imposed on the bulk inflow velocity in the $x$-direction.
The flame is initialized by the unstretched laminar flame solution, with the flame front laid in the middle of the $x$-direction. A stable, linear velocity forcing~\cite{Carroll_2013} is adopted to maintain the turbulent intensity from $x = 0.5 L$ to $9 L$ along the streamwise direction. The setup and accuracy of the DNS solver have been validated in Ref.~\cite{You_2019}.

\section{Results and discussions}
\label{sec:results}

\subsection{Turbulent burning velocity and flame area}
\label{sec:speed}

We define the turbulent burning velocity by the consumption speed
\begin{equation}\label{eq:sT}
  s_T = \dfrac{1}{\rho_u Y_{\mathrm{F},u}A_L\left(t_2-t_1\right)}\int_{t_1}^{t_2}\int_\Omega-\dot{\omega}_\mathrm{F} dVdt
\end{equation}
where the subscript $u$ is for unburnt conditions, $A_L=L_y\times L_z$ is the laminar flame area, $t_2-t_1$ is the period for collecting statistics, $\dot{\omega}_\mathrm{F}$ is the fuel consumption rate, and $\Omega$ denotes the entire computational domain. The turbulent flame area $A_T$ is evaluated at the isothermal surface of $T=T_\mathrm{peak}^\mathrm{F}$ via the marching cubes algorithm.

\begin{figure}
  \centering
  \includegraphics[width=67 mm]{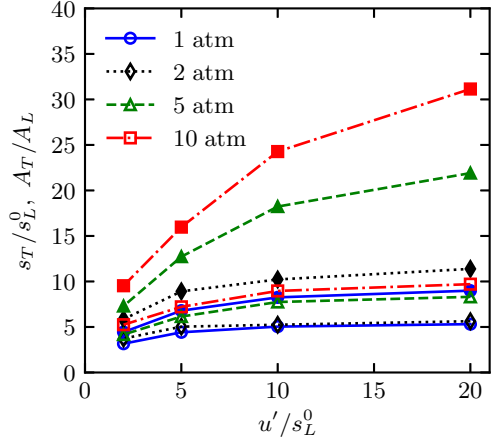}
  \caption{Normalized turbulent burning velocities (solid symbols) and flame areas (open symbols) at a range of pressures and turbulence intensities.}
  \label{fig:sTAT}
\end{figure}

Figure~\ref{fig:sTAT} plots ratios of burning velocities and flame areas against $u^\prime/s_L^0$ for different pressures.
We observe the bending of $s_T/s_L^0$ for a range of pressures. Meanwhile, the starting point of the bending occurs at larger $u^\prime/s_L^0$ with the increase of pressure, from $u^\prime/s_L^0\approx 5$ for $p=1\;\mathrm{atm}$ to $u^\prime/s_L^0\approx10$ for $p=10\;\mathrm{atm}$.
The diminished bending effect at elevated pressures was also reported in experiments for various fuels~\cite{Daniele_2011,Kobayashi_2013,Venkateswaran_2015}.
On the other hand, the bending of $A_T/A_L$ appears to be consistent at different pressures.
Thus the turbulent burning velocity is significantly accelerated at high pressures, e.g., $s_T/s_L^0$ at 10 atm is about four times of that at 1 atm in the present flame with $u'/s_L^0 = 20$.
By contrast, the variation of $A_T/A_L$ with the pressure is much less than that of $s_T/s_L^0$, e.g., $A_T/A_L$ at 10 atm is less than two times of that at 1 atm for $u'/s_L^0 = 20$.

\begin{figure}[htbp]
  \centering
  \includegraphics[width=67 mm]{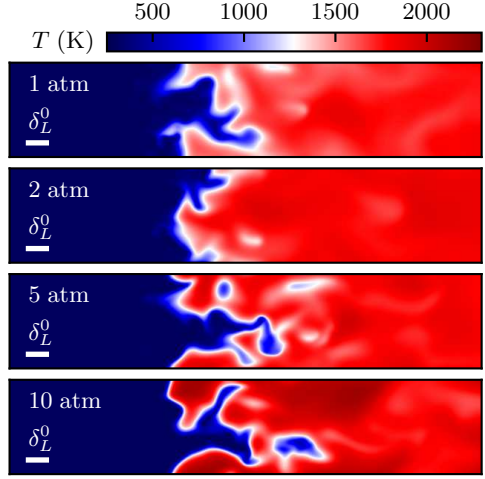}
  \caption{Temperature contours of the turbulent premixed flames at $u^\prime/s_L^0=10$ and various pressures.}
  \label{fig:contours}
\end{figure}

Figure~\ref{fig:contours} depicts temperature contours at various pressures and $u^\prime/s_L^0=10$, and flame shapes appear to be similar.
We also find that distributions of the normalized mean curvature of flames are self-similar at different pressures (not shown), as reported in previous studies~\cite{Savard_2017, Wang_2018}.
Additionally, the temperature around the flame front at 10 atm is much higher than that at 1 atm.

The Damkh\"oler assumption $s_T/s_L \sim A_T/A_L$ implies that the growth of $s_T$ is primarily due to the enhancement of $A_T$ by straining motions in low-intensity turbulence. The underlying assumption is that $s_L$ remains valid for the local propagation speed of flame fronts, and typically is taken as $s_L^0$ of unstretched laminar flames. With the consumption-based definition~\cite{Driscoll_2008}, the two ratios are linked by the stretch factor $I_0 = \langle s_L\rangle/s_L^0$~\cite{Bray_1991} as
\begin{equation}\label{eq:sTsL}
  \dfrac{s_T}{s_L^0} = I_0\dfrac{A_T}{A_L},
\end{equation}
where $\langle \cdot \rangle$ denotes the average over the turbulent flame front. This expression distinguishes contributions to $s_T$ into the area ratio from turbulence effects and the stretch factor due to the flame response under flow variations.
Therefore, Eq.~\eqref{eq:sTsL} and the different bending trends of $s_T/s_L^0$ and $A_T/A_L$ in Fig.~\ref{fig:sTAT} suggest that $I_0$ should depend on both turbulent intensity and pressure.


\subsection{Stretch factor}
\label{sec:stretch}

The stretch factor is close to unity at standard pressure, but it can vary at elevated pressures.
Thus we study the effects of turbulence intensity and pressure on $I_0$ in detail, along with the investigation on flame stretching and its influence on local flame speed.

The stretch factor
\begin{equation}\label{eq:I0sd}
  I_0 = \dfrac{\langle s_L\rangle}{s_L^0} \approx \dfrac{\langle s_d\rangle}{s_d^0}
\end{equation}
is estimated using the local displacement speed \cite{Uranakara_2016}
%
\begin{equation}
  s_d = \dfrac{\nabla\cdot\left(\lambda T\right)-\sum_{i=1}^{n_s}c_{p,i}\bm{j}_i\cdot\nabla T+c_p\dot{\omega}_T}{\rho c_p\lvert\nabla T\rvert}
\end{equation}
calculated on the isothermal surface of $T=T_\mathrm{peak}^\mathrm{F}$, where $\lambda$ is the thermal conductivity, $c_p$ and $c_{p,i}$ are respectively heat capacities of mixture and species $i$, $\bm{j}_i$ is the diffusion flux of species $i$, $n_s$ is the number of species, and $\dot{\omega}_T$ is the thermal production term.
Then Eq.~\eqref{eq:sTsL} is approximated by
\begin{equation}\label{eq:sTsL_approx}
  \dfrac{s_T}{s_L^0} \approx \dfrac{\langle s_d\rangle}{s_d^0}\dfrac{A_T}{A_L}.
\end{equation}

To compare different stretch effects, the stretch Karlovitz number $\mathrm{Ka_S} = \mathrm{Ka_T}+\mathrm{Ka_C}$ is decomposed into $\mathrm{Ka_T} = (\delta_L^0/s_d^0)a_t$ for tangential straining and $\mathrm{Ka_C} = (\delta_L^0/s_d^0)s_d\kappa$ for curvature stretch~\cite{Chen_1998}, where $a_t=\nabla\cdot\bm{u}-\mathbf{nn:}\nabla\bm{u}$ is the tangential strain rate, and $\kappa=\nabla\cdot\mathbf{n}$ is the mean curvature with the surface normal $\mathbf{n}=-\nabla T/\lvert\nabla T\rvert$. All these quantities are first calculated in $\Omega$ and then interpolated to the flame surface~\cite{Day_2009}.

\begin{figure*}
  \centering
  \includegraphics[width=137 mm]{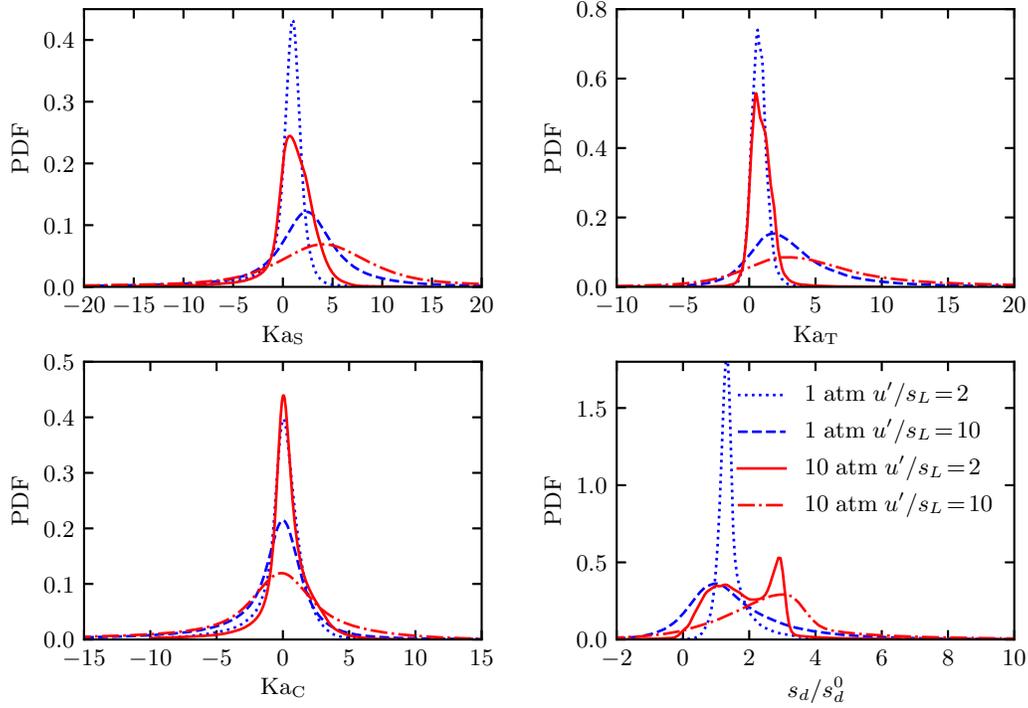}
  \caption{PDFs of the stretch-related Karlovitz numbers and local displacement speed at various pressures and turbulent intensities.}
  \label{fig:pdf_p}
\end{figure*}

Figure~\ref{fig:pdf_p} compares probability density functions (PDFs) of $\mathrm{Ka_S}$, $\mathrm{Ka_T}$, $\mathrm{Ka_C}$, and $s_d/s_d^0$ between pressures 1 atm and 10 atm with $u^\prime/s_L^0 = 2$ and 10. 
We observe that PDFs of both $\mathrm{Ka_S}$ and $\mathrm{Ka_T}$ in all the cases exhibit positive means, whereas PDFs of $\mathrm{Ka_C}$ keep roughly symmetric with the zero mean. 
%
Furthermore, PDFs of all the stretch-related $\mathrm{Ka}$ are widen with $u'/s_L^0$~\cite{Savard_2017} and pressure, and pressure effects on the PDFs become notable at high turbulence intensities, consistent with the different bending trends of $s_T/s_L^0$ in Fig.~\ref{fig:sTAT}.
It is clear that mean values of $\mathrm{Ka_S}$ and $\mathrm{Ka_T}$ increase with pressure for large $u'/s_L^0$, whereas the increase is slight for small $u'/s_L^0$.
The PDFs in Fig.~\ref{fig:pdf_p} indicate that $\mathrm{Ka_T}$ plays a dominant role in the positive stretching of turbulent flames, and $\mathrm{Ka_C}$ mainly contributes to broadening distributions of $\mathrm{Ka_S}$.

In Fig.~\ref{fig:pdf_p}d, PDFs of $s_d/s_d^0$ show pressure effects on the local flame propagation.
At $p=1$ atm, the distribution of $s_d/s_d^0$ is broadened with the turbulence intensity.
At $p=10$ atm, we observe two peaks in the PDF of $s_d/s_d^0$ for $u^\prime/s_L^0 = 2$.
The primary one is at $s_d/s_d^0\approx 3$, and the secondary one is close to the peak at $p=1$ atm around $s_d/s_d^0=1.1$. The two peaks are gradually merged with increasing $u^\prime/s_L^0$.

\begin{figure}
  \centering
  \includegraphics[width= 67mm]{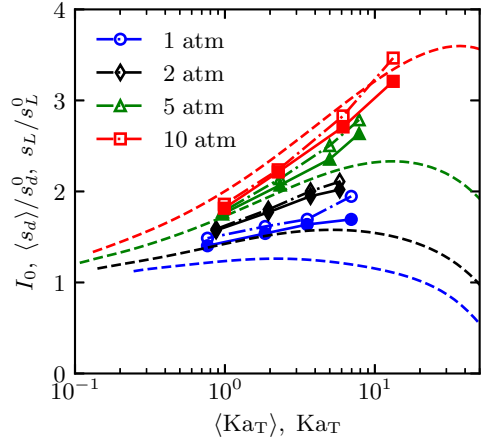}
  \caption{The stretch factor and its approximations in turbulent premixed flames and counterflow laminar flames (solid lines with solid symbols: $I_0$; dash-dotted lines with open symbols: $\langle s_d\rangle/s_d^0$; dashed lines: $s_L/s_L^0$). Both $I_0$ and $\langle s_d\rangle/s_d^0$ are plotted with respect to $\langle\mathrm{Ka_T}\rangle$, and $s_L/s_L^0$ is plotted against $\mathrm{Ka_T}$.}
  \label{fig:I0}
\end{figure}

Figure~\ref{fig:I0} plots the stretch factor $I_0=\left(s_T/s_L^0\right)/\left(A_T/A_L\right)$ in Eq.~\eqref{eq:sTsL} and its approximation $\langle s_d\rangle/s_d^0$ in Eq.~\eqref{eq:I0sd} with respect to $\langle\mathrm{Ka_T}\rangle$ from combustion DNS, and the normalized laminar flame speed $s_L/s_L^0$ against $\mathrm{Ka_T}$ from laminar counterflow flames.
%
%
The overall good agreement of $\langle s_d\rangle/s_d^0$ and $I_0$ in all the cases validates the approximation in Eq.~\eqref{eq:I0sd}.  
%

In laminar counterflow flames, $s_L/s_L^0$ increases with stretching owing to the negative Markstein number ($\mathrm{Ma}$) for the lean H$_2$/air mixture.
With the increase of pressure, $\mathrm{Ma}$ decreases due to the higher activation energy and Zel\textquotesingle dovich number.
Extensive studies of laminar flames \cite{Law_2006} have shown that $s_L/s_L^0$ grows with $\mathrm{Ka_T}$ for weak and moderate stretch. Moreover, we observe that the growth of $s_L/s_L^0$ is generally enhanced by pressure with negative $\mathrm{Ma}$.
Similarly in all the turbulence cases, $I_0$ and $\langle s_d\rangle/s_d^0$ also grow with pressure and $\langle\mathrm{Ka_T}\rangle$ which is proportional to the turbulence intensity.
The consistency between $\langle s_d\rangle/s_d^0$ and $s_L/s_L^0$ inspires us to model $I_0$ in turbulent flames using laminar flame results.




\subsection{Modeling of $s_T$ at high pressures}

We propose a predictive model for $s_T$ at a range of pressures and turbulent intensities. Based on the validation of Eq.~\eqref{eq:sTsL_approx}, the stretch factor and the flame area ratio are modeled separately.

For modeling the stretch factor, Fig.~\ref{fig:I0} suggests the consistency of $I_0$ and the response of $s_L/s_L^0$ to stretch of laminar flames, so we estimate $I_0$ using a lookup table $\mathcal{F}$ formed by laminar flame data on $s_L/s_L^0$ versus $\mathrm{Ka_T}$ as
\begin{equation}\label{eq:I0_model}
  I_0\left(\dfrac{u'}{s_L^0},p\right) = \dfrac{s_L}{s_L^0} = \mathcal{F}\left(\sqrt{\dfrac{p}{20 p_0}}\dfrac{u'}{s_L^0}\right),
\end{equation}
where an empirical relation $\mathrm{Ka_T}\approx\langle\mathrm{Ka_T}\rangle\approx\sqrt{p/20p_0}\left(u'/s_L^0\right)$ with $p_0=1$ atm is employed for incorporating pressure effects.

For modeling the flame surface ratio, we extend the $s_T$ model for standard pressure to high pressures.
The original model is given in Eq.~(3.30) in Ref.~\cite{You_2019} with a detailed derivation based on Lagrangian statistics of propagating surfaces, and it has been validated using combustion DNS with various fuels at $p=1$ atm.

We extend this model by including the influence of $I_0$ on $A_T/A_L$ at high pressures as
\begin{equation}\label{eq:AT_model}
  \dfrac{A_T}{A_L} = \exp\left\{T_\infty^*\left(\mathcal{A}+\mathcal{B}s_{L0}^0 I_0^2\right)\left[1-\exp\left(-\dfrac{\mathcal{C}\,\mathrm{Re}^{-1/4}}{T_\infty^*\left(\mathcal{A}+\mathcal{B}s_{L0}^0 I_0^2\right)}\dfrac{u^\prime}{s_L^0 I_0}\right)\right]\right\}.
\end{equation}
Here, universal model constants $\mathcal{A}=0.317$, $\mathcal{B}=0.033$, and $T_\infty^*=5.5$ for turbulence effects are obtained from Lagrangian statistics of propagating or material surfaces in non-reacting HIT, and $s_{L0}^0 = s_L^0/s_{L,\textrm{ref}}$ is a dimensionless laminar flame speed normalized by a reference value $s_{L,\textrm{ref}} = 1$ m/s. These model constants are the same as those in the original model~\cite{You_2019}.

In the improved model in Eq.~\eqref{eq:AT_model}, the unstretched laminar flame speed in the original one is replaced by $s_L^0 I_0$ with flame stretch effects.
The adapted model coefficient $\mathcal{C} = \mathcal{C}_0 I_0(u'/s_L^0 = 2,p)$ characterizes the combustion chemistry effect on the growth of $s_T/s_L^0$ in weak turbulence, where the constant $\mathcal{C}_0 = 2.5$ is suggested for hydrogen fuels in Ref.~\cite{You_2019}, and $I_0(u'/s_L^0 = 2,p)$ is used for recovering the linear growth of $s_T/s_L^0  = I_0 + \mathcal{C}_0 (u'/s_L^0)$ in weak turbulence with $u'/s_L^0 < 2$ from Taylor expansions of Eq.~\eqref{eq:AT_model}.
Furthermore, $\mathcal{A}+\mathcal{B}s_L$ in the original model approximates the growth rate of flame areas under the assumption of the constant local flame speed.
Considering the effect of flame stretch on the mean and variance of local flame speed in Fig.~\ref{fig:pdf_p}d, $s_L$ is approximated by $s_L^0I_0^2$ here.


Finally, substituting Eqs.~\eqref{eq:I0_model} and \eqref{eq:AT_model} and all the model coefficients into Eq.~\eqref{eq:sTsL} yields the $s_T$ model including pressure effects as
\begin{equation}\label{eq:sT_model}
  \begin{split}
  \dfrac{s_T}{s_L^0} = \mathcal{F}&\left(\sqrt{\dfrac{p}{20 p_0}}\dfrac{u'}{s_L^0}\right)\times \\
  \exp&\left\{\left(1.742+0.182s_{L0}^0 \mathcal{F}^2\right)\left[1-\exp\left(-\dfrac{2.5\mathcal{F}(\sqrt{p/(5p_0)})\,\mathrm{Re}^{-1/4}}{\left(1.742+0.182s_{L0}^0 \mathcal{F}^2\right)\mathcal{F}}\dfrac{u^\prime}{s_L^0}\right)\right]\right\}.
  \end{split}
\end{equation}
We remark that Eq.~\eqref{eq:sT_model} only depends on given flame/flow parameters, universal model constants, and laminar flame data, so it is a predictive model of $s_T$ for turbulent premixed flames.

\begin{figure}
  \centering
  \includegraphics[width=67 mm]{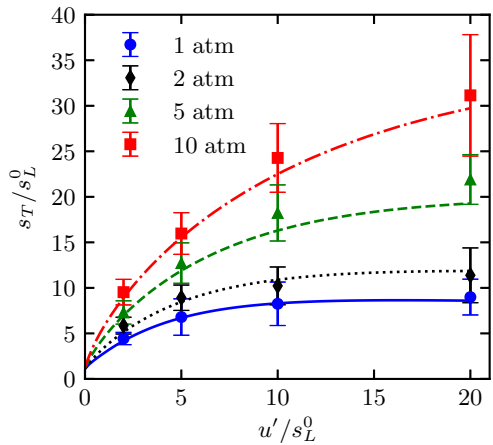}
  \caption{Comparison of $s_T$ calculated from combustion DNS (symbols with error bars for one standard deviation) and the proposed model Eq.~\eqref{eq:sT_model} (lines) at various pressures.}
  \label{fig:sT_model}
\end{figure}

The $s_T$ model in Eq.~\eqref{eq:sT_model} is validated by DNS results in Fig.~\ref{fig:sT_model}. In general, the present model shows quantitatively good predictions on important phenomena in turbulent premixed flames at a broad range of turbulent intensities and pressures.
Around 1 atm, the model predicts the bending curve at moderate and large $u'/s_L^0$.
As the pressure increases, our model well captures the rise of $s_T/s_L^0$ and the suppression of bending by accounting for pressure effects.

\section{Conclusions}

We elucidate the pressure effects on $s_T$ and propose a predictive $s_T$ model for turbulent premixed flames.
First we carried out a series of DNS for lean H$_2$/air turbulent premixed flames in HIT at $p=1$ to 10 atm and $u'/s_L^0=2$ to 20. The DNS results show bending curves of $s_T/s_L^0$ with respect to $u'/s_L^0$. In particular, $s_T/s_L^0$ for the same $u'/s_L^0$ increases with pressure, and the bending is suppressed at elevated pressures.

We further demonstrate both the turbulent flame area and stretch factor contribute to the rise of $s_T/s_L^0$ at elevated pressures for lean H$_2$/air flames.
The turbulent flame area has similar bending phenomena at various pressures.
The increased flame stretch sensitivity with pressure leads to the growth of $I_0$ at the high turbulence intensity and pressure.
In addition, large $I_0$ enhances the growth of $A_T$ via the increase of the local flame speed. 

From the analysis on DNS results, we extend the $s_T$ model for standard pressure \cite{You_2019} to high pressures via modeled $I_0$.
We find that the variation of $I_0$ with $\langle\mathrm{Ka_T}\rangle$ in turbulent premixed flames is similar to that of the laminar flame speed with flame stretch, so a lookup table for modeling $I_0$ is employed using laminar flame data at different pressures.
The flame area model based on Lagrangian statistics of propagating surfaces is extended by incorporating $I_0$ for the increased mean and variance of the local flame speed at high pressures.

Finally, we estimate $s_T/s_L^0$ as the product of modeled $I_0$ and $A_T/A_L$. This predictive $s_T$ model only depends on given flame and flow parameters, universal model constants, and laminar flame data.
The comparison between model predictions and DNS results shows overall good agreement, including the bending trends, in all the DNS cases at a broad range of turbulent intensities and pressures.


It is noted that the present study only considers the unburnt mixture with negative Markstein numbers in HIT, and the unity ratio of the turbulence length scale and flame thickness.
In the future work, $s_T$ for different fuels and flame geometries at elevated pressures will be investigated with further validations of the proposed $s_T$ model.
%


\section*{Acknowledgments}

We gratefully acknowledge Caltech, the University of Colorado at Boulder and Stanford University for licensing the NGA code used in this work. This work has been supported in part by the National Natural Science Foundation of China (Grant Nos. 91541204, 11925201, and 91841302).

\bibliography{ST_pressure_PCI} 

\begin{thebibliography}{10}
\expandafter\ifx\csname url\endcsname\relax
  \def\url#1{\texttt{#1}}\fi
\expandafter\ifx\csname urlprefix\endcsname\relax\def\urlprefix{URL }\fi
\expandafter\ifx\csname href\endcsname\relax
  \def\href#1#2{#2} \def\path#1{#1}\fi

\bibitem{Peters_2000}
N.~Peters, Turbulent Combustion, Cambridge University Press, 2000.

\bibitem{Lipatnikov_2002}
A.~Lipatnikov, J.~Chomiak, Turbulent flame speed and thickness: Phenomenology,
  evaluation, and application in multi-dimensional simulations, Prog. Energy
  Combust. Sci. 28 (2002) 1 -- 74.

\bibitem{Driscoll_2008}
J.~F. Driscoll, Turbulent premixed combustion: Flamelet structure and its
  effect on turbulent burning velocities, Prog. Energy Combust. Sci. 34 (2008)
  91 -- 134.

\bibitem{Daniele_2011}
S.~Daniele, P.~Jansohn, J.~Mantzaras, K.~Boulouchos, Turbulent flame speed for
  syngas at gas turbine relevant conditions, Proc. Combust. Inst. 33 (2011)
  2937 -- 2944.

\bibitem{Kobayashi_2013}
H.~Kobayashi, Y.~Otawara, J.~Wang, F.~Matsuno, Y.~Ogami, M.~Okuyama, T.~Kudo,
  S.~Kadowaki, Turbulent premixed flame characteristics of a
  {CO}/{H}$_2$/{O}$_2$ mixture highly diluted with {CO}$_2$ in a high-pressure
  environment, Proc. Combust. Inst. 34 (2013) 1437 -- 1445.

\bibitem{Bradley_2013}
D.~Bradley, M.~Lawes, K.~Liu, M.~Mansour, Measurements and correlations of
  turbulent burning velocities over wide ranges of fuels and elevated
  pressures, Proc. Combust. Inst. 34 (2013) 1519 -- 1526.

\bibitem{Venkateswaran_2015}
P.~Venkateswaran, A.~Marshall, J.~Seitzman, T.~Lieuwen, {Scaling turbulent
  flame speeds of negative Markstein length fuel blends using leading points
  concepts}, Combust. Flame 162 (2015) 375 -- 387.

\bibitem{Fragner_2015}
R.~Fragner, F.~Halter, N.~Mazellier, C.~Chauveau, I.~G{\"o}kalp, Investigation
  of pressure effects on the small scale wrinkling of turbulent premixed
  {Bunsen} flames, Proc. Combust. Inst. 35 (2015) 1527 -- 1535.

\bibitem{Wang_2015}
J.~Wang, S.~Yu, M.~Zhang, W.~Jin, Z.~Huang, S.~Chen, H.~Kobayashi, Burning
  velocity and statistical flame front structure of turbulent premixed flames
  at high pressure up to {1.0MPa}, Exp. Thermal Fluid Sci. 68 (2015) 196 --
  204.

\bibitem{Nguyen_2019}
M.~T. Nguyen, D.~W. Yu, S.~S. Shy, General correlations of high pressure
  turbulent burning velocities with the consideration of {Lewis} number effect,
  Proc. Combust. Inst. 37 (2019) 2391 -- 2398.

\bibitem{Kobayashi_2002_B}
H.~Kobayashi, T.~Kawahata, K.~Seyama, T.~Fujimari, J.-S. Kim, Relationship
  between the smallest scale of flame wrinkles and turbulence characteristics
  of high-pressure, high-temperature turbulent premixed flames, Proc. Combust.
  Inst. 29 (2002) 1793 -- 1800.

\bibitem{Im_2016}
H.~G. Im, P.~G. Arias, S.~Chaudhuri, H.~A. Uranakara, Direct numerical
  simulations of statistically stationary turbulent premixed flames, Combust.
  Sci. Technol. 188 (2016) 1182--1198.

\bibitem{Day_2009}
M.~S. Day, J.~B. Bell, P.-T. Bremer, V.~Pascucci, V.~Beckner, M.~Lijewski,
  Turbulence effects on cellular burning structures in lean premixed hydrogen
  flames, Combust. Flame 156 (2009) 1035 -- 1045.

\bibitem{Aspden_2017}
A.~J. Aspden, J.~B. Bell, M.~S. Day, F.~N. Egolfopoulos, Turbulence--flame
  interactions in lean premixed dodecane flames, Proc. Combust. Inst. 36 (2017)
  2005 -- 2016.

\bibitem{Uranakara_2016}
H.~A. Uranakara, S.~Chaudhuri, H.~L. Dave, P.~G. Arias, H.~G. Im, A flame
  particle tracking analysis of turbulence--chemistry interaction in
  hydrogen--air premixed flames, Combust. Flame 163 (2016) 220 -- 240.

\bibitem{Chen_1998}
J.~H. Chen, H.~G. Im, Correlation of flame speed with stretch in turbulent
  premixed methane/air flames, Proc. Combust. Inst. 27 (1998) 819 -- 826.

\bibitem{Nivarti_2017}
G.~Nivarti, S.~Cant, Direct numerical simulation of the bending effect in
  turbulent premixed flames, Proc. Combust. Inst. 36 (2017) 1903 -- 1910.

\bibitem{You_2019}
J.~You, Y.~Yang, Modelling of the turbulent burning velocity based on
  {Lagrangian} statistics of propagating surfaces, submitted to J. Fluid Mech.
  (2019).
\newblock \href {http://arxiv.org/abs/1911.00220} {\path{arXiv:1911.00220}}.

\bibitem{Savard_2017}
B.~Savard, S.~Lapointe, A.~Teodorczyk, Numerical investigation of the effect of
  pressure on heat release rate in iso-octane premixed turbulent flames under
  conditions relevant to {SI} engines, Proc. Combust. Inst. 36 (2017) 3543 --
  3549.

\bibitem{Wang_2017}
Z.~Wang, V.~Magi, J.~Abraham, Turbulent flame speed dependencies in lean
  methane-air mixtures under engine relevant conditions, Combust. Flame 180
  (2017) 53 -- 62.

\bibitem{Wang_2018}
X.~Wang, T.~Jin, Y.~Xie, K.~H. Luo, Pressure effects on flame structures and
  chemical pathways for lean premixed turbulent {H}$_2$/air flames:
  Three-dimensional {DNS} studies, Fuel 215 (2018) 320 -- 329.

\bibitem{Savard_2019}
B.~Savard, H.~Wang, A.~Wehrfritz, E.~R. Hawkes, Direct numerical simulations of
  rich premixed turbulent n-dodecane/air flames at diesel engine conditions,
  Proc. Combust. Inst. 37 (2019) 4655 -- 4662.

\bibitem{Venkateswaran_2011}
P.~Venkateswaran, A.~Marshall, D.~H. Shin, D.~Noble, J.~Seitzman, T.~Lieuwen,
  Measurements and analysis of turbulent consumption speeds of {H}$_2$/{CO}
  mixtures, Combust. Flame 158 (2011) 1602 -- 1614.

\bibitem{Vervisch_1995}
L.~Vervisch, E.~Bidaux, K.~N.~C. Bray, W.~Kollmann, Surface density function in
  premixed turbulent combustion modeling, similarities between probability
  density function and flame surface approaches, Phys. Fluids 7 (1995)
  2496--2503.

\bibitem{Girimaji_1992}
S.~S. Girimaji, S.~B. Pope, {Propagating surfaces in isotropic turbulence}, J.
  Fluid Mech. 234 (1992) 247--277.

\bibitem{Zheng_2017}
T.~Zheng, J.~You, Y.~Yang, Principal curvatures and area ratio of propagating
  surfaces in isotropic turbulence, Phys. Rev. Fluids 2 (2017) 103201.

\bibitem{Desjardins_2008}
O.~Desjardins, G.~Blanquart, G.~Balarac, H.~Pitsch, High order conservative
  finite difference scheme for variable density low {Mach} number turbulent
  flows, J. Comput. Phys. 227 (2008) 7125 -- 7159.

\bibitem{Ren_2008}
Z.~Ren, S.~B. Pope, Second-order splitting schemes for a class of reactive
  systems, J. Comput. Phys. 227 (2008) 8165--8176.

\bibitem{Pierce_2001}
C.~D. Pierce, Progress-variable approach for large-eddy simulation of turbulent
  combustion, Ph.D. thesis, Stanford University, Standford, CA, USA (2001).

\bibitem{Herrmann_2006}
M.~Herrmann, G.~Blanquart, V.~Raman, Flux corrected finite volume scheme for
  preserving scalar boundedness in reacting large-eddy simulations, AIAA J. 44
  (2006) 2879--2886.

\bibitem{Brown_1989}
P.~N. Brown, G.~D. Byrne, A.~C. Hindmarsh, {VODE}: A variable-coefficient {ODE}
  solver, SIAM J. Sci. Stat. Comput. 10 (1989) 1038--1051.

\bibitem{Li_2004}
J.~Li, Z.~Zhao, A.~Kazakov, F.~L. Dryer, An updated comprehensive kinetic model
  of hydrogen combustion, Int. J. Chem. Kinet. 36 (2004) 556--575.

\bibitem{Burali_2016}
N.~Burali, S.~Lapointe, B.~Bobbitt, G.~Blanquart, Y.~Xuan, Assessment of the
  constant non-unity {Lewis} number assumption in chemically-reacting flows,
  Combust. Theory Model. 20 (2016) 632--657.

\bibitem{Pope_2000}
S.~B. Pope, Turbulent Flows, Cambridge University Press, 2000.

\bibitem{Carroll_2013}
P.~L. Carroll, G.~Blanquart, A proposed modification to {Lundgren}'s physical
  space velocity forcing method for isotropic turbulence, Phys. Fluids 25
  (2013) 105114.

\bibitem{Bray_1991}
K.~N.~C. Bray, R.~S. Cant, Some applications of {Kolmogorov}'s turbulence
  research in the field of combustion, Proc. R. Soc. London A 434 (1991)
  217--240.

\bibitem{Law_2006}
C.~K. Law, {Combustion Physics}, Cambridge University Press, 2006.

\end{thebibliography}
\bibliographystyle{elsarticle-num}


\end{document}